\documentclass[12pt]{article}
\usepackage{amsmath}
\usepackage{amssymb}
\usepackage{graphics}
\usepackage{times}
\begin{document}

\begin{center}
{\LARGE \textbf{Esquisse d'une Synth\`ese\vspace{4pt}}}

{\large George A.J. Sparling\\\vspace{4pt} Laboratory of Axiomatics \\5814 Elgin Street\\\vspace{6pt} Pittsburgh, Pennsylvania, 15206, USA\bigskip }
\vspace{-10pt}
\begin{quote}\textbf{
\noindent \\
\noindent Since antiquity, from Euclid of Alexandria to Galileo Galilei to Immanuel Kant to Hermann Minkowksi to  Albert Einstein, the question of the nature of space and time has occupied scientists and philosophers \cite{euc1}-\cite{ein1}. In the four-dimensional space-time of Einstein's wonderful theory of gravity, the squared interval, in units such that the speed of light is unity, is the difference between a squared time increment and the sum of three squares representing the three dimensions of spatial change.  More recently higher dimensional theories have been proposed, which aim to unify gravity with the other forces in nature.  Such theories typically have a hyperbolic character in that there is one time variable and many spatial variables (rather than just three) in the formula for the squared interval \cite{all1}-\cite{rs2}.  Here a new physical theory is advanced, based on spinors, which clearly predicts that the basic extra dimensions are timelike: in its simplest form, there are three timelike and three spatial degrees of freedom.  It is expected that devices such as the Large Hadron Collider will be sensitive to these new degrees of freedom and thus one may hope that in the near future, this issue can be settled experimentally.  }\end{quote}\end{center}
In attempting to construct a theory, one sometimes has very little to go on.  One has to take strands of thought from many different disciplines and one has to try to weave them into a coherent whole, a concinnity.   One also has to be prepared to make major conceptual adjustments on the fly, as one brings in newer seminal ideas.    One should be attuned to the efforts of others and try to incorporate the essence of their best ideas, even if, in the end, one goes in a slightly different direction.  One should wield William of Ockham's Razor, but with parsimony \cite{moab0}!
\eject\noindent
Consider the simple act of taking a pencil and throwing it, spinning, into the air, so that it rotates completely around three times, before catching it again.  Common sense suggests that, \emph{ceteris paribus}, the pencil looks exactly the same at the end of this experiment as it did at the beginning.  However, we now know that the pencil is composed mostly of a variety of fermions and that under an odd number of complete rotations, amazingly, the wave function of each of these fermions \emph{changes sign}.  Only if we rerun our experiment,  can we be sure to restore the pencil to its pristine state.\\\\
Mathematically, this behaviour depends on the fact that the rotation group in three dimensions is not simply connected, but has a simply connected double cover, the group $\mathbb{SU}(2, \mathbb{C})$ of two by two unitary matrices, with complex entries, of unit determinant, whose topology is that of the real three-sphere, $\mathbb{S}^3$.  The lift of a single complete rotation to the group $\mathbb{SU}(2, \mathbb{C})$ is a curve connecting its identity element to its negative.  A second complete rotation is then required to return to the identity.  Quantities that transform with respect to the  group $\mathbb{SU}(2, \mathbb{C})$ are called spinors; in particular most known elementary particles are spinorial; these somehow transcend space and time. 
\\\\The purpose of the present work is to present the outline of a new physical theory, a significant extension of my earlier work with my former students Devendra Kapadia, Dana Mihai and Philip Tillman \cite{moab11}-\cite{moab6}.   I have had the remarkable good fortune to live long enough to solve a problem posed to me by Sir Roger Penrose, which has occupied me for the past forty years:  find a non-local spinorial approach to physics. \\\\  Until now, our previous work, although spinorial, appeared to lack the desired non-local feature.  Although, \emph{a priori},  the theory revolved around a variation on the concept of triality due to  Elie Cartan, the focus was more on the aspects of the theory coming from my own speciality: the twistor theory of Penrose,  Roy Kerr, Ivor Robinson, Ted Newman, Sir Michael Atiyah and  a small band of others \cite{ba1}-\cite{pen1}.  I realize now that that was a subtle philosophical error with deleterious consequences.  \\\\The correct ontology, I now believe, is one that is present in many philosophies from the earliest times: it is the trinity, three entities, which come together harmoniously forming the concinnity.  Here the three entities are initially conceptualized as space-time, twistor space and dual twistor space. 
\eject\noindent 
In the work "A primordial theory", relevant \emph{geometrical} and \emph{algebraic} ideas were developed, the principal objects being the exceptional algebra of 27 dimensions of Pascual Jordan,  associated to the split octaves, and the associated 56 dimensional phase space of Hans Freudenthal \cite{fr1}, \cite{jo1}.  I was shocked to discover in May of this year that there was a vast \emph{analytical} component that I had previously completely overlooked.  Perhaps the most surprising new feature is that the analytical structure, although originally developed in the context of conformally flat space-times, readily generalizes to curved space-time, where it gives a powerful new tool for deconstructing space-time, that is essentially spinorial in a non-local and deep way, thereby achieving one of the original aims of the twistor program.  All previous essays along these lines have languished, although it is important to notice that Lane Hughston had, for the case of flat space, a similar approach \cite{hu1}.\vspace{-10pt}
\subsection*{The $\Xi$-transform}
The analytic structure in question is a \emph{transform}, which I will call the $\Xi$-transform.  It is perhaps most easily expressed using the two-component complex spinor formalism for relativity.  The constituents of the transform are as follows:
\begin{itemize} \item The space-time $\mathbb{M}$, which is a smooth real manifold of dimension four.  Its phase space is the co-tangent bundle $\mathbb{T^*M}$ of $\mathbb{M}$, consisting of all pairs $(x, p)$ with $x$ in $\mathbb{M}$ and $p$ a co-vector at $x$.  Denote by $\alpha = \theta.p$, the contact one-form on $\mathbb{T^*M}$; here $\theta$ is the vector-valued canonical one-form of $\mathbb{M}$ and the dot denotes the dual pairing of a vector with a co-vector.
\item A Lorentzian metric $g$ for $\mathbb{M}$, of signature $(1, 3)$, such that $\mathbb{M}$ is space and time orientable and has a chosen spin structure.
\item The co-spin bundle $\mathbb{S}^*$ is the set of all pairs $(x, \pi)$, where $\pi$ is a primed co-spinor at the point $x$ in $\mathbb{M}$; so $\mathbb{S}^*$ is a complex vector bundle of two complex dimensions over $\mathbb{M}$ (so as a real vector bundle $\mathbb{S}^*$ has four dimensional fibers).  Recall that $\mathbb{S}^*$ is equipped with a complex symplectic form $\epsilon$, a global section of the exterior product of  $\mathbb{S}^*$, with itself, such that $g = \epsilon\otimes \overline{\epsilon}$.   
\item We assume given the spin connection of the type of Tullio Levi-Civita and Jan Schouten, denoted $d$, which is torsion free and annihilates $\epsilon$ and $g$. 
\item Each co-spinor $\pi$ gives rise to a co-vector $p_\pi =  \pi\otimes\overline{\pi}$, which is zero if $\pi$ is zero and is otherwise null and future-pointing.  So $\mathbb{S}^*$ behaves like a "square root" of the bundle of future-pointing null co-tangent vectors.  Denote by $P: \mathbb{S}^* \rightarrow \mathbb{T^*M}$ the map which takes $(x, \pi) \in \mathbb{S}^*$ to $(x, p_{\pi})\in \mathbb{T^*M}$.
\end{itemize}
Note that $p_{t\pi} = |t|^2 p_{\pi}$, for any spinor $\pi$ and any complex number $t$, so that $p_{\pi}$ is insensitive to the overall phase of the spinor $\pi$.  Very  few constructions directly depend on this phase, the tensor $\mathcal{F}$, described later, being the most crucial.  We henceforth delete the zero section from $\mathbb{S}^*$, so all spinors are taken to be non-zero. 
\begin{itemize} \item The bundle of co-spin frames, $\mathbb{B}$ consists of all ordered triples $(x, \pi_+, \pi_-)$, where $x$ is in $\mathbb{M}$ and the $\pi_\pm$ are primed co-spinors at $x$, which are normalized against each other by the equation $2\pi_+\wedge \pi_- = \epsilon$. Note that $\mathbb{B}$ is a principal $\mathbb{SL}(2, \mathbb{C})$-bundle over $\mathbb{M}$.
\item Note also that if the non-zero co-spinor $\pi_-$ is given at  $x \in \mathbb{M}$, the space of all normalized pairs $(\pi_+, \pi_-)$ at  $x \in \mathbb{M}$ is a one-dimensional complex affine space, so has two real dimensions, since if $(\pi_0, \pi_-)$ is normalized, for some co-spinor $\pi_0$, then the general normalized pair $(\pi_+, \pi_-)$ can be written $(\pi_0 + \lambda \pi_-, \pi_-)$, with $\lambda$ an arbitrary complex number.
\item There are natural maps $\Pi_{\pm}: \mathbb{B} \rightarrow \mathbb{S}^*$, which map $(x, \pi_+, \pi_-) \in \mathbb{B}$ to $(x, \pi_{\pm}) \in \mathbb{S}^*$.  
\item The (future-pointing) null geodesic spray on the null co-tangent bundle lifts naturally to the space $\mathbb{S}^*$ to give a vector field denoted $\mathcal{N}$.  The trajectories of $\mathcal{N}$ represent an affinely parametrized future-pointing null geodesic on $\mathbb{M}$, together with a co-spinor $\pi$, parallelly propagated along the geodesic, such that $g^{-1}(p_\pi)$ is the (normalized) tangent vector to the geodesic.  
\item A quantity on $\mathbb{S}^*$ that is invariant along the vector field $\mathcal{N}$ is called a twistor quantity.  In particular, a function $f(x, \pi)$ killed by $\mathcal{N}$ is called a twistor function.  Such a function is said to be real homogeneous of (integral) degree $k$ in $\pi$, if $f(x, t\pi) = t^k f(x, \pi)$, for any non-zero \emph{real} number $t$.  If $f$ is complex-valued, we say that  $f(x, \pi)$ is complex homogeneous of degree $k$ in $\pi$, if $f(x, t\pi) = t^k f(x, \pi)$, for any non-zero \emph{complex} number $t$.  Then the real homogeneous twistor functions depend on six real variables, whereas the complex homogeneous twistor functions depend on five real variables.   
\item The space $\mathbb{B}$ naturally carries \emph{two} horizontal vector fields, denoted $\mathcal{N}_{\pm}$, corresponding to the (horizontal) lifts of the vector field $\mathcal{N}$ along the maps  $\Pi_\pm$. The trajectories of these vector fields represent an affinely parametrized null geodesic in space-time, together with a normalized spin-frame $(\pi_+, \pi_-)$, parallelly propagated along the null geodesic, such that for $\mathcal{N}_\pm$, the vector $g^{-1}(p_{\pi_\pm})$ is a normalized tangent vector to the null geodesic.
\end{itemize} 
\eject\noindent
Now we can proceed to the $\Xi$-transform: 
\begin{itemize} \item Let $\beta$ be a given differential three-form on the space $\mathbb{S}^*$.
\item Let $\gamma$ be a given horizontal future-pointing null geodesic curve (i.e. an integral curve of the vector field $\mathcal{N}$)  in $\mathbb{S}^*$.  
\item Pull $\beta$ back to the space $\mathbb{B}$ along the natural projection $\Pi_+$ to give the three-form $\beta_+ = \Pi_+^*(\beta)$ on the space $\mathbb{B}$.  
\item Let $\gamma_- = \Pi_-^{-1}(\gamma)$ denote the inverse image under the map $\Pi_-$ of the horizontal curve $\gamma$.  We think of the space $\gamma_-$ as a "fattened" version of the null geodesic $\gamma$, in that a two-real-dimensional affine space is located at each point of the null geodesic, rather than just the spinor $\pi_-$ itself.  In particular $\gamma_-$ has real dimension three.
\item  Integrate the three-form $\beta_+$ over the space $\gamma_-$ to give a number, denoted $\Xi(\beta)(\gamma)$.  As the curve $\gamma$ varies, we get, by definition, the $\Xi$-transform $\Xi(\beta)$ of the three-form $\beta$ as a function on the space $\mathbb{S}^*$, which is invariant along the null geodesic spray $\mathcal{N}$, so $\Xi(\beta)$ is a twistor function.  
\end{itemize} 
This completes the general description of the $\Xi$-transform.
\\\\
The special case that is relevant for the remainder of this work is the case that $\beta =  if(x, \pi) \epsilon^{-1}(\pi, d\pi) \overline{\epsilon}^{-1}(\overline{\pi}, d\overline{\pi})\theta.p_\pi$, where $f(x, \pi)$ is a real-valued twistor function, real homogeneous of degree minus four in the variable $\pi$.   For this case, we write the transform as $f \rightarrow \Xi(f)$.   Then one can show that $\Xi(f)$ is itself a real twistor function, this time real homogeneous of degree minus two and that the transform is conformally invariant.  Note that $\beta$ can be written also as the multiplication of the function $f$ by the pull back to $\mathbb{S}^*$ of the three-form $\frac{1}{2}\omega(p, g(\theta), dp, dp)$ where $\omega$ is the contravariant alternating orientation tensor associated to the metric $g$ and $dp$ is the tautological co-vector valued one form on $\mathbb{T^*M}$ that incorporates the Levi-Civita connection of $\mathbb{M}$.  \\\\Summarizing, this key particular case of the $\Xi$-transform gives a conformally invariant operator taking twistor functions of degree minus four to twistor functions of degree minus two.  In particular, both the input and output functions are functions with six real degrees of freedom.
\eject\noindent
Let us now specialize to the conformally flat case.   To understand this fully, let us temporarily call the transform we have just constructed $\Xi_1$ and let us introduce two other transforms, denoted $\Xi_2$ and $\Xi_3$, at first sight unrelated to $\Xi_1$.
\begin{itemize} \item The second transform is given by the following integral formula:
\[  \Xi_2(f)(g, h) = \int_{p \in \mathbb{G}}  f(p, g^{-1} ph) \omega_p.\]
Here $g$, $h$ and $p$ belong to a compact Lie group, $\mathbb{G}$, $\omega_p$ is Haar measure for $\mathbb{G}$ and $f$ is a smooth function on $\mathbb{G}\times \mathbb{G}$.  The integral is taken over all $p \in \mathbb{G}$.  For the present purposes we specialize to the case that $\mathbb{G} = \mathbb{SU}(2, \mathbb{C})$.  Then $\mathbb{G}$ is topologically a real three-sphere, $\mathbb{S}^3$, so $\Xi_2$ maps functions of six real variables, specifically functions on the product space $\mathbb{S}^3 \times \mathbb{S}^3$, to themselves.
\end{itemize} 
The third transform explicitly uses $\mathbb{O}(4, 4)$-triality, which we briefly recall in outline \cite{moab11, ba1}.  It involves three real eight-dimensional vector spaces $\mathbb{A}$, $\mathbb{B}$ and $\mathbb{C}$, say, each equipped with an $\mathbb{O}(4, 4)$ dot product,  together with a certain real trilinear form mapping $\mathbb{A} \times \mathbb{B} \times \mathbb{C}$ to the reals, denoted by $(xyz)$, for $(x, y, z) \in  \mathbb{A} \times \mathbb{B} \times \mathbb{C}$. Dualizing this trilinear form gives rise to three real bilinear  maps $\mathbb{A}\times \mathbb{B} \rightarrow \mathbb{C}$,  $\mathbb{B}\times \mathbb{C}  \rightarrow \mathbb{A}$ and $\mathbb{C}\times \mathbb{A}\rightarrow \mathbb{B}$, denoted by parentheses, such that, for example, $((xy)x) = x.x y$, and $(xy).z = (zx).y = (yz).x = (xyz)$, for any $x$, $y$ and $z$ in $\mathbb{A}$, $\mathbb{B}$ and $\mathbb{C}$,  where the dot product is the appropriate $\mathbb{O}(4, 4)$ inner product.   The whole theory is then symmetrical under permutations of the three  vector spaces.  We say that $x$ in $\mathbb{A}$ and $y$ in $\mathbb{B}$ are \emph{incident} if they are both non-zero and yet $(xy) = 0$; this entails that $x$ and $y$ are both null.  Further, given a null $y \ne 0$ in $\mathbb{B}$, the space of all $x$ in $\mathbb{A}$, such that $(xy) = 0$ is a real, totally null, self-dual, four-dimensional vector space.
\begin{itemize} \item The third transform now proceeds as follows.  Let $f(x)$ be a smooth real-valued function homogeneous of degree minus four, defined for all non-zero null $x\in \mathbb{A}$.  Then $f(x)\hspace{3pt}  x \wedge dx\wedge dx \wedge dx$ is a closed three-form on the null cone of $\mathbb{A}$ taking values in $\Omega^4(\mathbb{A})$, the fourth exterior product of $\mathbb{A}$ with itself.  Then define a function $\xi_3(f)$, on the null cone of $\mathbb{B}$, taking values in $\Omega^4(\mathbb{A})$, by the formula, valid for any null vector $y \ne 0$ in $\mathbb{B}$:
\[ \xi_3(f)(y) = \int_{x\hspace{3pt} \textrm{incident with} \hspace{3pt} y}  f(x) \hspace{3pt} x \wedge dx\wedge dx \wedge dx.\]
Here the integral is taken over the natural homology three-sphere in the complement of the origin of the space of all $x$ such that $(xy) = 0$.  \end{itemize}
\eject\noindent
For this transform, there is a beautiful additional subtlety.   First one shows that the output takes values in the self-dual part of $\Omega^4(\mathbb{A})$.   Next one observes that $\Omega^4(\mathbb{A})$ has real dimension $70$, so the self-dual part has real dimension $35$.   But this is exactly the real dimension of symmetric trace-free  tensors of valence two in $\mathbb{B}$ and one shows that there is a natural isomorphism between the two spaces.  For $y \in \mathbb{B}$, which is null, the tensor $y \otimes y$ is symmetric and trace-free.  Let $\epsilon(y \otimes y) \in \Omega^4(\mathbb{A})$ denote the (self-dual) image of $y\otimes y$ under this isomorphism.  Then one shows that the output $\xi_3(f)(y)$ naturally \emph{factorizes}: \[ \xi_3(f)(y) = \Xi_3(f)(y) \epsilon(y \otimes y).\]   Since $\xi_3(f)(y)$ is, from its definition, homogeneous of degree zero in $y$, it follows that $\Xi_3(f)$ is a real-valued function on the space of all null non-zero vectors $y$ in $\mathbb{B}$, homogeneous of degree minus two in $y$.\\\\
Note that there is one such transform for each ordered pair from the set $\{\mathbb{A}, \mathbb{B}, \mathbb{C}\}$, giving \emph{six} such transforms in all (three of these initially take values in self-dual forms and the other three take values in anti-self-dual forms).
\\\\
We now have the following results:
\begin{itemize} \item The transforms $\Xi_1$, $\Xi_2$ and $\Xi_3$ \emph{coincide, mutatis mutandis}.  This means that one can prove results for one of the transforms and deduce analogous results for the others, which might be harder to get at directly.  
\item  For the transform $\Xi_2$, we introduce the Casimir operator $\mathcal{C} = C_1 - C_2$, where each of $C_1$ and $C_2$ is the Casimir operator of $\mathbb{SU}(2, \mathbb{C})$ acting on the first and second factors of the product $\mathbb{SU}(2, \mathbb{C})\times \mathbb{SU}(2, \mathbb{C})$, respectively.  So $\mathcal{C}$ is a differential operator of the second order.   Then we have the beautiful result that $\mathcal{C}\circ\Xi_2 = \Xi_2\circ \mathcal{C} = 0$.  Equivalently, the kernel of $\Xi_2$ contains the image of $\mathcal{C}$ and vice-versa.  It is probably true that the kernel of $\Xi_2$ \emph{exactly matches} the image of $\mathcal{C}$ and vice-versa, but at the time of writing, this has only been proved fully under the restriction that the functions involved are finite sums of spherical harmonics.  Note that $\mathcal{C}$ is the (ultra-hyperbolic) wave operator in six dimensions for the natural metric on $\mathbb{S}^3 \times \mathbb{S}^3$ of signature $(3, 3)$.
\eject\noindent
\item Using our translation principle, one can re-formulate these results at the level of the other operators $\Xi_1$ and $\Xi_3$.     For $\Xi_3$, the space of all null non-zero vectors $y$ in $\mathbb{B}$, where we identify $y$ with $ty$, for $t > 0$, is a space of topology $\mathbb{S}^3 \times \mathbb{S}^3$, which now has only a (natural) conformal structure, rather than a fixed metric structure, as in the case of $\Xi_2$.   At this point the work of C. Robin Graham, Ralph Jenne, Lionel Mason and myself comes into play: it shows (amongst other results) that there is a natural conformally invariant second order wave operator, denoted $\square$, for conformally flat geometries in six-dimensions, which maps  functions of conformal weight minus two to functions of conformal weight minus four \cite{graham1}.   These are precisely the weights for the transform $\Xi_3$ and the analogous relation for the operator $\Xi_3$ is then $\square\circ \Xi_3 = \Xi_3\circ \square = 0$.
\item   Finally for the spin bundle, the analogous operator can be defined as follows.  Let $f(x, \pi)$ be a given twistor function, of degree minus two.  Then, using abstract spinor and vector indices, since $f$ is constant along the null geodesic spray, we may write its gradient, $\partial_a f$, with respect to the variable $x$, as $\partial_a f = \pi_{A'} \overline{f}_A + \overline{\pi}_A f_{A'}$, where $f_{A'}$ is of degree minus three.  Denote by $\partial^{A'}$ the (complex) gradient with respect to the spinor $\pi_{A'}$.    Then we write $ \square(f) = i(\partial^{B'} f_{B'} - \overline{\partial}^B \overline{f}_B)$.  Then it can be shown that $\square(f)$ is a twistor function of degree minus four.  Now our basic result is: $\square\circ \Xi_2 = \Xi_2\circ \square = 0$.
\end{itemize}  
What can one say in curved space-time?  It will probably take many years and a major research program to fully reveal the structure of the $\Xi$-transform.   For a moment, let us dwell on the stumbling blocks that prevented progress from the direction of twistor theory, in the past.   Twistor theory worked beautifully in the cases of self-dual gauge fields and self-dual gravity and related equations \cite{maswood1}.  It seems clear, in retrospect, that the reason for this success is that these systems of equations were integrable; the methods of the theory always used this fact, implicitly or explicitly.    Nevertheless, for ordinary non-self-dual gravity, there was some success: the $\mathcal{H}$-space (self-dual) theory of Newman and Penrose, although valid only for analytic space-times, arises out of the gravitational radiation data of a real non-self-dual space-time \cite{new1}, \cite{pen6}.  Also the equations of Frederick Ernst for stationary axi-symmetric space-times were shown by Richard Ward to admit a twistor interpretation \cite{ward1}. However it is believed that neither the source-free gauge-field equations, nor the vacuum equations of gravity are integrable in general.  For example it is believed that the so-called solutions of the ninth type of Luigi Bianchi exhibit chaos.  Also chaos seems to appear near a generic singularity \cite{bkl1} . \\\\
For the first time the $\Xi$-transform appears likely to give a precise criterion sorting out the more tractable space-times from the rest, according to the nature of its image.  We say that the space-time is \emph{coherent}, if and only if the image of the $\Xi$-transform obeys a \emph{pseudo-differential} equation.  If not, we say the space-time is chaotic.  Then, in this language, we have shown that conformally flat space-time is coherent.  One would conjecture that all the real space-times, that have in the past proved to be amenable to twistor-type treatments, are coherent.  As of the time of writing we have been able to show by direct calculation that the prototypical space-times of Devendra Kapadia and myself are coherent, at least for complex homogeneous input functions, giving the first known example of a curved space-time that is such \cite{moab3}.  Note that this classification is inherently \emph{non-perturbative} and gives a \emph{co-ordinate independent} definition of dynamical chaos.\vspace{-11pt}
\subsection*{Going up to six dimensions}
\vspace{-3pt}
At the level of space-time, there is a disparity of dimensions: space-time is four dimensional, whereas our twistor spaces depend on functions of six real variables.   The key question now arises: is there a realm in physics in which the triality is more manifest?   This would require enlarging space-time from four to six real dimensions.  We would like to do this in a natural way building directly from the conventional space-time theory.  Very remarkably, it emerges that we can!  \\\\There are two clues:  first consider the situation in conformally flat space-time.   The basic spaces of the triality have the symmetry group $\mathbb{O}(4, 4)$.   However for the twistor spaces to relate to ordinary physics, this group is too big.  We want the twistor spaces to reproduce the standard successful quantization of massless particles, using (holomorphic) sheaf cohomology, due to Hughston, Penrose and myself \cite{pen5}.  In particular, we want to implement the standard twistor commutation relations, which form the algebra of Werner Heisenberg: $[Z^a, Z^b] = i \hbar \omega^{ab}$; here the indices run from $1$ to $8$.  This entails that we need a symplectic form $\omega^{ab}$; in fact, to recover the standard theory, we easily see that $\omega^{ab}$ must be a complex-structure for the eight-dimensional vector space, such that its symmetry group is reduced from $\mathbb{O}(4, 4)$ to the group $\mathbb{U}(2, 2)$.   A similar story applies to the other twistor space.   However, for the space-time triality space, we need to separate out the space-time:  this entails reducing the symmetry group from $\mathbb{O}(4, 4)$ to $\mathbb{O}(4, 2)$.  Note that this is essentially distinct from the reductions for the twistor spaces: although the groups $\mathbb{SO}(4, 2)$ and $\mathbb{SU}(2, 2)$ are locally isomorphic, the latter being a double cover of the former, they sit inside the group $\mathbb{O}(4, 4)$ in \emph{different} places. 
\eject\noindent
Surprisingly, it emerges that \emph{a single technique} does the job simultaneously for all three triality spaces.   For the case of the triality space that one wants to be space-time, say the space $\mathbb{A}$, one simply selects an oriented two-dimensional subspace $\mathbb{J}$ of the eight-dimensional vector space with a positive definite induced metric.  Then the orthogonal subspace is six dimensions, which intersects the null cone of the triality space in a five dimensional space, whose real projectivization gives the four-dimensional space-time, conformally compactified, with a natural conformal structure and the correct conformal symmetry group.  Let $j_1$ and $j_2$ be unit orthogonal elements in the subspace $\mathbb{J}$, such that $\{j_1, j_2\}$ is an oriented basis for $\mathbb{J}$.   For $b \in \mathbb{B}$ and for $c \in \mathbb{C}$, denote by $J(b)$ in $\mathbb{B}$ and $K(c)$ in $\mathbb{C}$, the quantities $J(b) = (j_1(j_2b))$ and $K(c) = (j_2(j_1c))$, respectively.  Then it is easy to see from the properties of the triality that $J$ and $K$ are complex structures for the spaces $\mathbb{B}$ and $\mathbb{C}$, giving these spaces the desired reduction from $\mathbb{O}(4, 4)$ to $\mathbb{SU}(2, 2)$.  Also the structures $J$ and $K$ are invariant under rotations of the basis $\{j_1, j_2\}$.\\\\
The second clue comes from the structural spin tensor of the spin-bundle of space-time.  This takes the form $\mathcal{F} = i\theta^a\otimes (\overline{\pi}_A d\pi_{A'} - \pi_{A'} d\overline{\pi}_A)$.   It has three fundamental properties, which encode precise details of the space-time:  first its skew part gives the two-form used by Edward Witten in his argument for positive energy; second, properties of the exterior  derivative of the skew part can be used to analyze the Einstein vacuum equations; third its symmetric part, when restricted to any hypersurface, gives the conformal structure of the type of Charles Fefferman for the twistor theory of that hypersurface as shown by myself \cite{moab3, feff1, witten1}. In particular, it provides the \emph{central fact of twistor theory}, from which all else follows.  In moving to a higher dimensional framework, one would like to extend this tensor, to maintain that same control over the field equations and over the twistor theory.  \\\\
Remarkably, it emerges that in extending to six dimensions, with a conformal structure of signature $(3, 3)$, the tensor $\mathcal{F}$ has a beautiful, completely natural extension, which actually looks better than the original:  it is the tensor, still called $\mathcal{F}$, given by the formula: $\mathcal{F} = \theta^{\alpha\beta} \otimes \pi_\alpha d\pi_{\beta}$; note that $i$ does not appear.  Here $d$ represents the spin connection in six dimensions and we are using the fact that the spin group for the group $\mathbb{SO}(3, 3)$ is the group $\mathbb{SL}(4, \mathbb{R})$.   The basic spinor $\pi_\alpha$ is then \emph{four real dimensional}, carrying the fundamental (dual) representation of $\mathbb{SL}(4, \mathbb{R})$.  This means that the spinors \emph{restrict naturally}, without any loss of information, to four-dimensional submanifolds: the correspondence with the spinors of Richard Brauer and Hermann Weyl  is just $\pi_\alpha \rightarrow (\pi_{A'}, \overline{\pi}_A)$ \cite{weyl1}.
\eject\noindent
The canonical one-form $\theta^{\alpha\beta}$ is skew, so has the required six degrees of freedom.   Decomposing into the spinors of relativity we get a quartet: $(\theta^{AB}, \theta_1^{AB'}, \theta_2^{A'B}, \theta^{A'B'})$.    Here $\theta^{AB} = \theta \epsilon^{AB}$ may be construed as giving a kind of complex "dilaton" field and has $\theta^{A'B'} = \overline{\theta} \epsilon^{A'B'}$ is its complex conjugate.  To recover the standard four-dimensional metric one would want the one-form $\theta$ to vanish on the four-manifold.  Then for the rest of the canonical one-form $\theta^{\alpha\beta}$ we have two relations $\overline{\theta_1^{AA'}} =  \theta_2^{A'A}$ and $\theta_1^{AA'} = - \theta_2^{A'A}$.  These relations, taken together, mean that $\theta_1^{AA'} = i \theta^{AA'} = - \theta_2^{A'A}$, where $\theta^{AA'}$ is self-conjugate, giving, on restriction, the required real canonical one-form of relativity.  Then $\mathcal{F} = \theta^{\alpha\beta} \otimes \pi_\alpha d\pi_{\beta}$ restricts to $i\theta^{AA'} \otimes (\overline{\pi}_A d\pi_{A'} -  \pi_{A'} d\overline{\pi}_A)$, exactly the Fefferman tensor, the necessary factors of $i$ emerging naturally, even though the spinors of the ambient space are entirely real.\\\\
However there is a subtle catch, which is where the two clues need to be brought to bear simultaneously.   When the ambient spin connection is restricted to the space-time submanifold, there is no reason that it should preserve the complex structure of the space-time spinors.  From the ambient point of view, if $D$ is the space-time spin connection, which \emph{does} preserve the complex structure, the restricted spin connection can read, for example: $d\pi_{A'} = D\pi_{A'} + \Gamma_{A'}^A \overline{\pi}_A$.   This is a disaster, since the field $\Gamma_{A'}^A$ is a vector-valued one-form, so has spin-two components, giving gravity extra spin-two degrees of freedom, that are probably unphysical.  \\\\The resolution is beautifully simple: one postulates that the conformal geometry has a conformal Killing vector, or if the actual metric is specified that it have  a Killing vector.   Recall that if a metric $g_{ab}$ has a Killing vector $t^a$, then the tensor $\partial_a t_b = F_{ab}$ is skew.   Here indices are abstract and $\partial_a$ is the Levi-Civita connection of $g_{ab}$ \cite{pen4}. Then a standard formula gives the covariant derivative of $F_{ab}$: $\partial_a F_{bc} = 2R_{bca}^{\hspace{12pt}d}t_d$.   Here $R_{bca}^{\hspace{12pt} d}$ is the Riemann tensor of $\partial_a$.   In particular, if the Killing vector vanishes on space-time, then the restriction of $F_{bc}$ is covariantly constant, so becomes part of the space-time structure.   So here we demand that the metric in six-dimensions have a Killing symmetry, whose orbits are circles, such that the space-time is the set of fixed points of the symmetry (one thinks of the symmetry as a rotation in the "two-plane" perpendicular to  the four-dimensional "axis").  Looking back at our construction in the conformally flat case, one sees that that is exactly what one has: the rotation is simply the ordinary rotation in the space $\mathbb{J}$, keeping the orthogonal space fixed: this "axis" then provides the space-time.  The derivative of the Killing field provides the invariant complex structure needed for the spinors and twistors in the space-time.\\\\
Thus we are suggesting  that space-time extends naturally and conformally into six-dimensions, where it is the set of fixed points of an appropriate conformal Killing vector field.  But the signature of the six dimensions is, quite unambiguously, $(3, 3)$.  So the extra dimensions are quite definitely timelike!    Note that we have effectively invoked here a philosophical principle, that in the context of physics, we may attribute to Paul Dirac: if it is elegant, then it must be right!  \cite{dirac1, dirac2}. This is perhaps the most dangerous principle in philosophy! \\\\Notice that our approach has three immediate pay-offs: first space-time is a kind of "brane", allowing the ideas of Joseph Polchinski to come into play \cite{pol1}.   Second we have a natural place for arguments of the type given by Lisa Randall and Raman Sundrum, who make the case that the extra dimensions can compensate for the apparent weakness of gravity \cite{rs1, rs2}.  Also if we factor out by the Killing field, we will have signature $(2, 3)$, giving a suitable arena to apply the ideas of Juan Maldacena \cite{mal1}.\\\\
Note that we do not necessarily require the full strength of the symmetry: it needs only to be asymptotically a symmetry as the space-time is approached.
\vspace{-10pt} \subsection*{The concinnity}
Finally we address the concinnity.  Here we are not yet in position to provide a definitive theory.   However there are some constraints:  
\begin{itemize} \item It must provide an arena for the fundamental fermionic quantum liquid of  Shou-Cheng Zhang and Jianping Hu \cite{zh1, zh2, moab9}.
\item It must be geometrical, analytical and algebraic (Hopf).
\item It must encode the concepts of sheaves and sheaf cohomology that are critical in twistor theory \cite{pen0, pen5, pen4}.
\item It must unify quantum mechanics and geometry.
\item It would be desirable that it include the main ideas of current physics, apart from those already mentioned.
\end{itemize}
\eject\noindent
For the last, I do not pretend to be an expert, but will proffer some ideas. The famous Calabi-Yau theory of Philip Candelas, Gary Horowitz, Andrew Strominger and Edward Witten, as described in my earlier work, seems to find a home in the null hypersurface twistor spaces, where the hypersurface has no vertex, but terminates in a singularity \cite{calabi1, yau1, yau2, moab4}.  So essentially that theory classifies the structure of space-time singularities.   Similarly the counting of black hole states works with horizons, which are null hypersurfaces \cite{host1}.  The manifolds of Dominic Joyce are more problematic, probably living outside the usual space-time arena and in our six-dimensional space, the Calabi-Yau theory in space-time being a limiting case \cite{joy1}.  Support here comes from the breakthrough work of Pawel Nurowksi and myself on the structure of third order differential equations \cite{nuro1}.  \\\\
The structure we need is so powerful that it must involve deep mathematics.  So I conjecture that it is a coherent topos (a generalized approach to sheaf theory due to F. William Lawvere and Myles Tierney), with a triangulated structure of the type developed by Jean Louis Verdier and Alexandre Grothendieck to provide the cohomology \cite{lawv1}.   Hints of the latter structure appear in the cohomological character of the $\Xi$-transform for conformally flat space-time found  above.   Where can we start to look for this structure in the space-time?  I believe it lies in the ensemble of all conformally invariant hyperbolic differential or even pseudo-differential operators on the space-time, together with the "modules" that they act on.  These somehow express the \emph{non-analytic essence of hyperbolicity}, which is the key new feature introduced into physics by James Clerk Maxwell and Albert Einstein.\\\\
The full structure will be a non-commutative geometry, more than likely a non-commutative string theory, as in the work of Alain Connes \cite{connes1}.  It should have the property that looked at ("observed!") one way, involving "going to the boundary", one recovers the basic quantum twistor space, describing massless particles, whereas looked at another such way, one recovers the relevant space-time phase space (the null co-tangent bundle).  Note that the very fact that there are twistor and space-time based descriptions of the same basic reality, that of massless particles, hints at a common ontology.  The structure, which would not be in itself dynamical, because of the lack of a preferred time concept, then creates the required dynamics at the "edge".  Then the fundamental  "seat of pants" picture of string theory may be recovered as a generalization of the $\Xi$-transform: a method of transferring information between the various edges.  However, unlike conventional string theory, where the strings at the boundary of the pants are much of a muchness, here the three boundary strings belong to \emph{three different spaces} \cite{wit2}. 
\eject\noindent
The concepts presented here should have analogues in other areas.  There may be a direct application in the context of superfluid helium three, which has a natural $\mathbb{SU}(2, \mathbb{C}) \times \mathbb{SU}(2, \mathbb{C})$ structure; if this pans out one may be able to test the present theory  using superfluids, and incorporate some of the ideas of Grigori Volovik \cite{volovik1, bain1}.  Finally there should be a close analogue for the theory of solitons, extending the deep recent work of Lionel Mason and Claude Le Brun and linking it with the ideas of Alexei Bondal and Dimitri Orlov \cite{mason5, bond1}.\\\\
This work is dedicated to the memory of my sister Fru.  I thank all those who have contributed to my ideas over the years.  I thank Sir Roger Penrose for being so inspirational and for giving me a chance and Sir Michael Atiyah for support at critical moments. I would like to thank my ancestors and my family, especially my mother, father, step-father, Erin, Zed/Zee and Camille.  Also my yoga teachers Adrienne, Alison and Saeeda.  Also my recent students David Hillman, Devendra Kapadia, Dana Mihai, Suresh Maran, Jocelyn Quaintance and  Philip Tillman.  Also my friends and colleagues Maciej Dunajski, J\"org Frauendiener, Lionel Mason and Pawel Nurowski.  Finally  I would like to thank the philosophers Alexander Afriat, Steve Awodey, Jonathan Bain and Rita Marija Malikonyte-Mockus.  I would also like to thank Alexander Afriat and the University of Urbino for inviting me to describe some of these results earlier this summer, which greatly helped me clarify my thoughts \cite{moab12}.   Finally I thank the Healey Foundation for financial support.

\end{document}